\documentclass[amssymb,pra,twocolumn,longbibliography,notitlepage,superscriptaddress]{revtex4-1}

\newcommand{\kets}[1]{| #1 \rangle}
\newcommand{\bras}[1]{\langle #1 |}

\usepackage[%
colorlinks=true,
urlcolor=blue,
linkcolor=blue,
citecolor=blue
]{hyperref}

\usepackage{color} 

\usepackage{amsmath}                          
\usepackage{amsfonts}                         
\usepackage{amssymb}                          
\usepackage{amsthm}                           
\usepackage{mathdots}                         
\usepackage{booktabs}
\usepackage[braket, qm]{qcircuit}   
\usepackage{algorithmicx} 
\usepackage{algorithm}
\usepackage{algpseudocode}

\usepackage{graphicx}                         
\usepackage{subfigure}                        
\usepackage{overpic}

\begin{document}  

\title{Hybrid quantum-classical algorithms for solving quantum chemistry in Hamiltonian-wavefunction space}
\author{Zhan-Hao Yuan}\affiliation{National Laboratory of Solid State Microstructures
and School of Physics, Nanjing University, Nanjing 210093, China}

\author{Tao Yin}
\email{tao.yin@artiste-qb.net}
\affiliation{Yuntao Quantum Technologies, Shenzhen, 518000, China}

\author{Dan-Bo Zhang}
\email{dbzhang@m.scnu.edu.cn}
\affiliation{Guangdong Provincial Key Laboratory of Quantum Engineering and Quantum Materials, GPETR Center for Quantum Precision Measurement and SPTE, South China Normal University, Guangzhou 510006, China}
\affiliation{Frontier Research Institute for Physics,
	South China Normal University, Guangzhou 510006, China}


%

\begin{abstract}
Variational quantum eigensolver~(VQE) typically optimizes variational parameters in a quantum circuit to prepare eigenstates for a quantum system. Its applications to many problems may involve a group of Hamiltonians, e.g., Hamiltonian of a molecule is a function of nuclear configurations. In this paper, we incorporate derivatives of Hamiltonian into VQE and develop some hybrid quantum-classical algorithms, which explores both Hamiltonian and wavefunction spaces for optimization. Aiming for solving quantum chemistry problems more efficiently, we first propose mutual gradient descent algorithm for geometry optimization by updating parameters of Hamiltonian and wavefunction alternatively, which shows a rapid convergence towards equilibrium structures of molecules. We then establish differential equations that governs how optimized variational parameters of wavefunction change with intrinsic parameters of the Hamiltonian, which can speed up calculation of energy potential surface. Our studies suggest a direction of hybrid quantum-classical algorithm for solving quantum systems more efficiently by considering spaces of both Hamiltonian and wavefunction. 
\end{abstract}

%

\maketitle
\section{Introduction} 
Variational quantum eigensolver opens a promising paradigm for solving eigenstates of Hamiltonians on near-term quantum processors with hybrid quantum-classical optimization~\cite{MYung2014,McClean2016,OMalley2016,Kandala2017,McArdle2019,JLiu2019}. It has received intensive studies since it provides a practical avenue to exploit the power of quantum computing for many fundamental problems, ranging from quantum chemistry~\cite{yung_14,mcclean_16,omalley_16,kandala_17,grimsley_19,arute2020hartreefock}, quantum many-body systems~\cite{liu_19,kokail_19,dallairedemers2020application}, and many other applications~\cite{anschuetz_18,xu2019variational,lubasch_20}. 
The original VQE is designed to solve the ground state for a single Hamiltonian, and variants of VQE have been developed for solving excited states~\cite{mcclean_17,nakanishi_19,Higgott_19,greenediniz2019generalized,zhang_20_vvqe}, finite-temperature quantum systems~\cite{wu_19,verdon_19,liu_19,chowdhury2020variational,wang_20,zhu2019generation}, and so on.  

Many practical problems may involve a group of Hamiltonians characterized by intrinsic parameters that describes the system, for instance, electronic Hamiltonian for a molecule is a  function of nuclear configurations. The energy dependence of nuclear configurations, namely potential energy curve/surface~(PEC/PES) with bond lengths and angles, account for many properties of chemical reaction such as transition states, reaction rate, etc~\cite{McArdle2020}. While it requires VQE for solving molecules under many different configurations and the computational cost can be large, it is possible to explore relations between Hamiltonians to give more efficient algorithm. For instance, collective optimization has been developed to update variational parameters of wavefunctions for different Hamiltonians jointly~\cite{Zhang2020}, guided by the continuousness of Hamiltonian. As such a simple strategy works, it is natural to exploit more concrete relations, such as first order derivative of Hamiltonian with bond lengths or angles. While calculations of energy derivatives have been proposed in Ref.~\cite{OBrien_2019, parrish_2019, mitarai_2020}, it is still lack of exploiting Hamiltonian derivatives for variational solving quantum chemistry in an enlarged space of Hamiltonians and wavefunctions, where both variational parameters of wavefunction and intrinsic parameters of Hamiltonians can be searched and optimized.  

In this paper, we incorporate Hamiltonian derivative in the framework of VQE, and develop hybrid quantum-classical algorithms for solving quantum chemistry in the space of both Hamiltonians and trial wavefunctions.  We first propose mutual gradient descent algorithm for geometry optimization, which aims to find equilibrium structure of molecules more efficiently. As optimized variational parameters of wavefunctions are expected to change continuously with varying Hamiltonians, we also establish differential equations that reveal the relation. With a set of differential equations and an initial condition, we can calculate the energy potential surface without using VQE for every configurations, including both ground state and excited states. 
By numeral simulations, we demonstrate the algorithms for some representative molecules. Our work 
suggests Hamiltonian derivative as an important ingredient in VQE for solving quantum chemistry or other physical systems involving a group of Hamiltonians.

The paper is organized as follows. In section.~\ref{sec:method}, we first introduce VQE and propose two new algorithms exploiting Hamiltonian derivatives. In section.~\ref{sec:result}, we apply those methods to molecules such as H$_2$, LiH and H$_4$ numerically. Finally, we give a summary of this work in section.~\ref{sec:summary}.

\section{Variational quantum eigensolvers with Hamiltonian derivative} \label{sec:method}
Variational quantum eigensolver aims to solve eigenstates for a given Hamiltonian. 
In quantum chemistry, the Hamiltonian depends on the nuclear configurations of the molecule, as positions of atoms are fixed by the Born-Oppenheimer approximation. Then, we can solve a molecule for a range of nuclear configurations, e.g., to get the energy potential surface. As Hamiltonian will change continuously with the configuration parameters, we can expect accessing the Hamiltonian derivative with nuclear configuration can be helpful for solving quantum chemistry problems more efficiently. We incorporate this idea into VQE into two different ways. First, we propose mutual gradient descent algorithm for geometry optimization, which finds the lowest-energy molecule structure for a molecule in an enlarged parameter space. Second, we establish differential equations that uncovers the relation between the optimized parameter of variational wavefunctions
and the Hamiltonians.  

\subsection{Variational quantum eigensolver}
We consider Hamiltonians of a molecule are $H(\lambda)$ with intrinsic parameter $\lambda$. The variational wavefunction ansatz is denoted as $\kets{\psi(\boldsymbol{\theta})}$, where $\boldsymbol{\theta}\in \mathcal{R}^K$. We also adopt a density matrix denotation $\psi(\boldsymbol{\theta})=\kets{\psi(\boldsymbol{\theta})}\bras{\psi(\boldsymbol{\theta})}$.
The Hamiltonian can be written as a summation of local Hamiltonians, 
\begin{equation}
H(\lambda)=\sum_{i=1}^{N} c_i(\lambda) L_i,
\end{equation}
where a local Hamiltonian $L_i$ can be written as product of Pauli matrices. We denote $\boldsymbol{c}(\lambda)=[c_1(\lambda),c_2(\lambda),...,c_N(\lambda)]^T$ and $\boldsymbol{L}=[L_1,L_2,...,L_N]^T$. Thus we can write $H(\lambda)=\boldsymbol{c}^T(\lambda)\boldsymbol{L}$. As $\boldsymbol{c}(\lambda)$ shall be continuous function of $\lambda$, derivative of $H(\lambda)$ is defined as $\frac{\partial\boldsymbol{c}^T(\lambda)}{\partial\lambda}\boldsymbol{L}$. 

For a single Hamiltonian with particular $\lambda$, the VQE works as follows. An variational ansatz $\kets{\psi(\boldsymbol{\theta})}=U(\boldsymbol{\theta})\kets{\psi_0}$ is used to parametrize a ground state. The initial state  $\kets{\psi_0}$  is usually choosen as a good classical approximation for the ground state of $H$. In quantum chemistry, for instance, $\kets{\psi_0}$ can be chosen as a Hartree-Fock state.  $U(\boldsymbol{\theta})$ is an unitary operator parameterized with $\boldsymbol{\theta}$, which can encode quantum correlation into the ground state. The essential task is to find parameters $\boldsymbol{\theta}$ that minimizes the energy
\begin{equation}\label{eq:ene}
\mathcal{E}(\boldsymbol{\theta};\lambda)=\bras{\psi(\boldsymbol{\theta})}H(\lambda)\kets{\psi(\boldsymbol{\theta})}=\text{Tr}(\psi(\boldsymbol{\theta})H(\lambda)).
\end{equation}

In the process of optimization, the quantum processor prepares $\psi(\boldsymbol{\theta})$ and performs measurements to evaluate $\mathcal{E}(\boldsymbol{\theta};\lambda)$, which can be reduced into 
\[
\mathcal{E}(\boldsymbol{\theta};\lambda)=\boldsymbol{c}^T(\lambda)\boldsymbol{\mathcal{L}}(\boldsymbol{\theta}).
\] 
where $\boldsymbol{\mathcal{L}}(\boldsymbol{\theta})=\text{Tr}(\psi(\boldsymbol{\theta})\boldsymbol{L})$.
Here, quantum average of each component of $\boldsymbol{L}$ corresponds to a joint measurement on multiple qubits that can be implemented on a quantum processor.
The classical computer updates parameters $\boldsymbol{\theta}$ according to received data from the quantum processor, e.g., using gradient descent 
\begin{equation}\label{eq:gd}
\boldsymbol{\theta}^{t}=\boldsymbol{\theta}^{t-1}-\eta_A \frac{\partial}{\partial{\boldsymbol{\theta}}}\mathcal{E}(\boldsymbol{\theta}^{t-1};\lambda),
\end{equation}
where $\eta_A$ is the step size. 
Calculating the gradients with respect to a target cost function (here is $\mathcal{E}(\boldsymbol{\theta};\lambda)$) which can be obtained with the same quantum circuit on a quantum processor, using the shift rule~\cite{JLi2017,Schuld2019} or numeral differential.  The optimization for energy minimization reaches a zero gradient descent 
\begin{equation}\label{eq:match_wf}
\frac{\partial\mathcal{E}(\boldsymbol{\theta};\lambda)}{\partial\boldsymbol{\theta}}\equiv\boldsymbol{c}^T(\lambda)\frac{\partial\boldsymbol{\mathcal{L}}(\boldsymbol{\theta})}{\partial\boldsymbol{\theta}}=0,
\end{equation}
which we may call it as wavefunction-matching condition. 

\subsection{Mutual gradient descent algorithm for geometry optimization}

Geometry optimization is an important task in  computational chemistry,  which is key to understand molecular structures and chemistry reactions. VQE has been applied for geometry optimization, which optimization along the path of energy potential surfaces~\cite{OBrien_2019}. In this work, we propose a more efficient hybrid algorithm by directly minimizing $\mathcal{E}(\boldsymbol{\theta};\lambda)$ in the enlarged parameter space $(\boldsymbol{\theta},\lambda)$, without referring to optimize VQE for each fixed $\lambda$.

We may divide the procedure into wavefunction-matching and Hamiltonian matching processes. The former is just VQE that optimizes $\boldsymbol{\theta}$ by fixing $\lambda$. The Hamiltonian matching can be formulated as follows. For a $\psi(\boldsymbol{\theta})$, to find a Hamiltonian in the group of Hamiltonians $H(\lambda)$, we propose the Hamiltonian matching condition,
\begin{equation}\label{eq:match_H}
\frac{\partial\mathcal{E}(\boldsymbol{\theta};\lambda)}{\partial\lambda}\equiv\frac{\partial\boldsymbol{c}^T(\lambda)}{\partial\lambda}\boldsymbol{\mathcal{L}}(\boldsymbol{\theta})=0.
\end{equation}
To reach this condition, $\lambda$ can be updated using gradient descent as following, 
\begin{equation}\label{eq:gd_lambda}
\lambda^{t}=\lambda^{t-1}-\eta_B \frac{\partial\mathcal{E}(\boldsymbol{\theta};\lambda^{t-1})}{\partial\lambda}.
\end{equation}
It should be noted that the term $\frac{\partial\boldsymbol{c}^T(\lambda)}{\partial\lambda}$ is given with classical computers using opensource packages~(OpenFermion~\cite{OpenFermion} or HiqFermion~\cite{Hiq}). Then, once $\boldsymbol{\mathcal{L}}(\boldsymbol{\theta})$ has been evaluated on a quantum process, the optimization with Eq.~\eqref{eq:gd_lambda} can be run iteratively on a classical computer. This is in contrast to the optimization process of Eq.~\eqref{eq:gd} where the quantum processor and the classical computer should be used repeatedly. For this reason, finding a best matched Hamiltonian is a task without using quantum resources.  


We propose mutual gradient descent for geometry optimization, which start from an initial
$\psi(\boldsymbol{\theta_0})$ for a given Hamiltonian $H(\lambda)$, and optimize $\boldsymbol{\theta}$ and $\boldsymbol{\lambda}$ alternatively. The algorithm is as follow (N, T are hyper-parameters controlling the iterative step):

\begin{algorithm} [H]
        \caption{Mutual Gradient Descent~(MGD)}  
        \begin{algorithmic}[1] 
            \Require $\psi(\boldsymbol{\theta^0})$, $\lambda^0$, $N$, $T$
            \Ensure $\lambda^t$
            \Function {MGD}{$\psi(\boldsymbol{\theta}^0), \lambda^0, N, T$}  
                \State $\lambda^t \gets \lambda^0$ 
                \State $\boldsymbol{\theta}^t \gets \boldsymbol{\theta}^0$
                \Repeat
                \For{$i = 0 \to N-1$}
                \State $\lambda^{t}=\lambda^{t-1}-\eta_B \frac{\partial\mathcal{E}(\boldsymbol{\theta};\lambda^{t-1})}{\partial\lambda} $
                \EndFor
                \For{$i = 0 \to T-1$}
                \State $\boldsymbol{\theta}^{t}=\boldsymbol{\theta}^{t-1}-\eta_A \frac{\partial}{\partial{\boldsymbol{\theta}}}\mathcal{E}(\boldsymbol{\theta}^{t-1};\lambda) $
                \EndFor
                \Until{Convergence}
                \EndFunction
        \end{algorithmic}  
\end{algorithm}  

\subsection{Differential equations for calculating energy potential surface}
We turn to solve the energy potential surface, which lies at the heart of quantum computational chemistry. The energy potential surface describes the dependence of ground or low-energy excited states with bond lengths and angles for a molecule. For VQE, it can be expected that optimized $\boldsymbol{\theta}^*$ vary continuously with intrinsic parameters $\lambda$ of the Hamiltonian, namely $\boldsymbol{\theta}^*(\lambda)$ is a function of $\lambda$. We can reveal explicitly this function, which can very useful for calculating energy potential surface. 

For this, we set $\lambda'=\lambda+\delta\lambda$ and $\boldsymbol{\theta}'=\boldsymbol{\theta}+\delta\boldsymbol{\theta}$ for the wavefuntion matching condition in Eq.~\eqref{eq:match_wf}, 
and then expand $\boldsymbol{c}^T(\lambda')\frac{\partial\boldsymbol{\mathcal{L}}(\boldsymbol{\theta'})}{\partial\boldsymbol{\theta'}}=0$ as (omitting second order terms)
\begin{equation}
(\boldsymbol{c}^T(\lambda)+\frac{\partial\boldsymbol{c}^T(\lambda)}{\partial\lambda}\delta\lambda)(\frac{\partial\boldsymbol{\mathcal{L}}(\boldsymbol{\theta})}{\partial\boldsymbol{\theta}}+\frac{\partial^2\boldsymbol{\mathcal{L}}(\boldsymbol{\theta})}{\partial\boldsymbol{\theta}^2}\delta\boldsymbol{\theta})=0.
\end{equation}
Then, one can get a differential equation, 
\begin{equation}\label{eq:df1}
\frac{\partial\boldsymbol{c}^T(\lambda)}{\partial\lambda}\frac{\partial\boldsymbol{\mathcal{L}}(\boldsymbol{\theta})}{\partial\boldsymbol{\theta}} +\boldsymbol{c}^T(\lambda) \frac{\partial^2\boldsymbol{\mathcal{L}}(\boldsymbol{\theta})}{\partial\boldsymbol{\theta}^2} \frac{d\boldsymbol{\theta}(\lambda)}{d\lambda}=0. 
\end{equation} 
The system of differential equations of Eq.~\eqref{eq:df1} contains a number of $N\times K$ equations. We can use a simple case $N=K=1$ to illuminate the meaning of this differential equation. Eq.~\ref{eq:df1} is simplified as 
\begin{equation}\label{eq:df1_simple}
\boldsymbol{c}(\lambda)^T\boldsymbol{\mathcal{L}}''(\theta)d\theta=-\boldsymbol{c}'(\lambda)^T\boldsymbol{\mathcal{L}}'(\theta)d\lambda.
\end{equation}

A more inspiring way is to write Eq.~\eqref{eq:df1_simple} as,
\begin{equation}\label{eq:balance}
\partial_\theta\partial_\theta\mathcal{E}(\theta;\lambda)d\theta(\lambda)=-\partial_\theta\partial_\lambda \mathcal{E}(\theta;\lambda)d\lambda
\end{equation}
We may take $\kappa_{\theta\lambda}=\partial_\theta\partial_\lambda\mathcal{E}(\theta;\lambda)$ and $\kappa_{\theta\theta}=\partial_\theta\partial_\theta\mathcal{E}(\theta;\lambda)$ as two elastic coefficients.  The numeric algorithm can be done as follows,
\begin{equation}
\theta_{i+1}^*=\theta_{i}^*-(\lambda_{i+1}-\lambda_i)\frac{\kappa_{\theta\lambda}}{\kappa_{\theta\theta}}.
\end{equation} 
Note that $\kappa_{\theta\lambda}$ and $\kappa_{\theta\theta}$ can be evaluated with a quantum computer~(by numeral differential or analytic differential using the shift rule). Calculating each $\theta_{i}^*$ may inevitably bring some errors. A slightly modification can be made by using gradient descent in Eq.~\eqref{eq:gd} to make sure an optimized $\theta_{i}^*$ is indeed obtained, which is expected to take very few steps.

Let's consider cases with higher dimensional parameters. For
case of $\boldsymbol{\theta}=(\theta_1,\theta_2)$, apply Eq.~\eqref{eq:df1} to Eq.~\eqref{eq:balance} similarly, we can write the differential equation as follow.
\begin{eqnarray} \label{eq:two_theta}
\partial_{\theta_1}\partial_\lambda \mathcal{E}(\boldsymbol{\theta};\lambda)d\lambda+\partial_{\theta_1}\partial_{\theta_1}\mathcal{E}(\boldsymbol{\theta};\lambda)d\theta_1+\partial_{\theta_1}\partial_{\theta_2}\mathcal{E}(\boldsymbol{\theta};\lambda)d\theta_2=0 \nonumber \\
\partial_{\theta_2}\partial_\lambda \mathcal{E}(\boldsymbol{\theta};\lambda)d\lambda+\partial_{\theta_2}\partial_{\theta_1}\mathcal{E}(\boldsymbol{\theta};\lambda)d\theta_1+\partial_{\theta_2}\partial_{\theta_2}\mathcal{E}(\boldsymbol{\theta};\lambda)d\theta_2=0 \nonumber \\
\end{eqnarray}
Apparently, the discussion above can be promoted to cases with any dimension, so a matrix form of the differential equation can be represented as 
\begin{equation}
\boldsymbol{A} d\boldsymbol{\theta} =- \boldsymbol{b}d\lambda,
\end{equation}
where

\begin{equation}
A =
\left[ \begin{array}{cccc}
\partial_{\theta_1}\partial_{\theta_1}\mathcal{E} & \partial_{\theta_1} \partial_{\theta_2}\mathcal{E}  & \cdots & \partial_{\theta_1} \partial_{\theta_n}\mathcal{E} \\
\partial_{\theta_2} \partial_{\theta_1}\mathcal{E} & \partial_{\theta_2} \partial_{\theta_2}\mathcal{E}  & \cdots & \partial_{\theta_2} \partial_{\theta_n}\mathcal{E}\\
\vdots & \vdots & \ddots & \vdots \\

\partial_{\theta_n} \partial_{\theta_1}\mathcal{E} & \partial_{\theta_n} \partial_{\theta_2}\mathcal{E}  & \cdots & \partial_{\theta_n} \partial_{\theta_n}\mathcal{E}

\end{array} 
\right ],
b=
\left[ \begin{array}{c}
\partial_{\theta_1} \partial_{\lambda}\mathcal{E}\\
\partial_{\theta_2} \partial_{\lambda}\mathcal{E}\\
\vdots\\
\partial_{\theta_n} \partial_{\lambda}\mathcal{E}
\end{array}
\right ]
\end{equation}

one can neatly write the numeral solver as, 
\begin{equation}
\boldsymbol{\theta}_{i+1}^*=\boldsymbol{\theta}_{i}^*-(\lambda_{i+1}-\lambda_i)\boldsymbol{A}^{-1}\boldsymbol{b}.
\end{equation}

It can be viewed as a numeral solution for the differential equation with a hybrid quantum-classical algorithm. More advanced numeral methods may be applied for improving the precision. 

The different equations can be readily for solving the energy potential surface with an initial condition, which can be obtained by using VQE for a molecule at a fixed configuration. This avoids to apply VQE for every configurations. Moreover, the different equations are in principle applicable for all eigenstates, which is determined by the initial condition. Thus, different equations can be easily incorporated into VQE for calculating EPS consisting of excited states~\cite{nakanishi_19,zhang_20_vvqe}. 

\begin{figure*}
	\subfigure[]{
	\begin{minipage}[t]{0.48\textwidth}
		\centering
		\includegraphics[width=\textwidth]{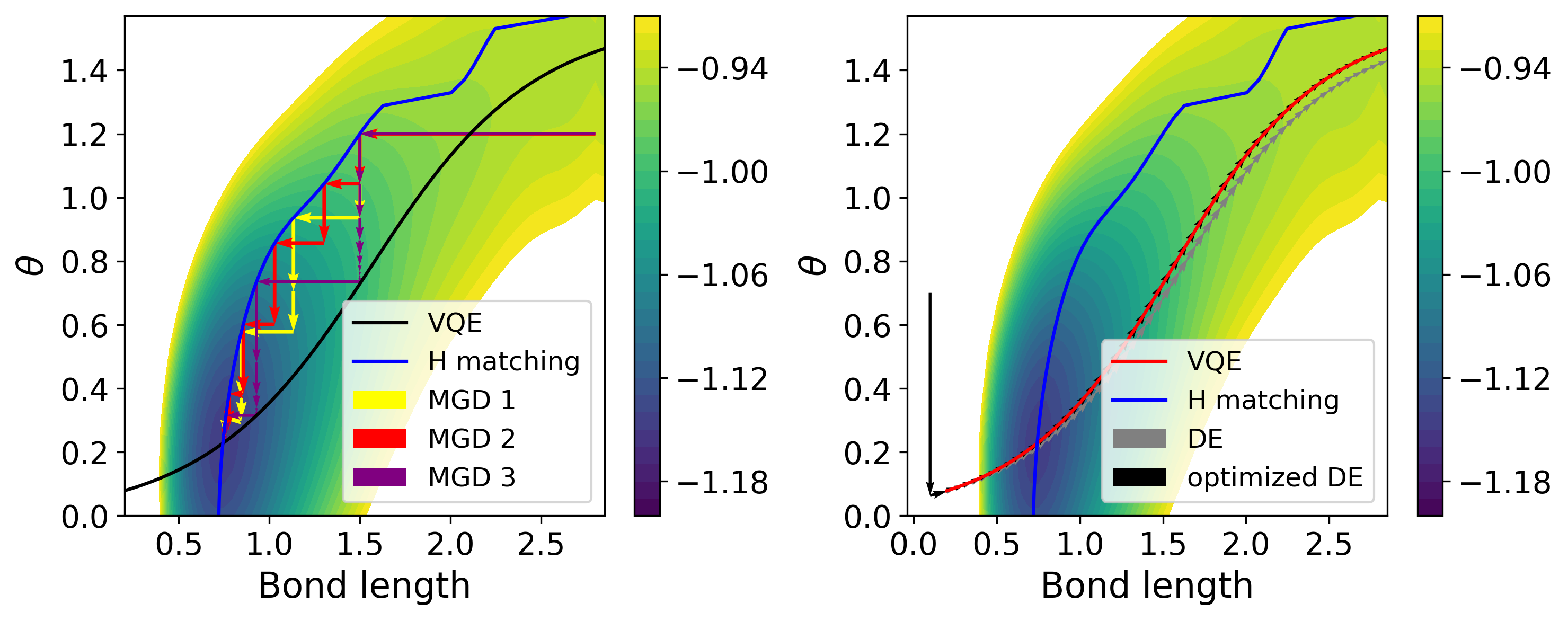}
	\end{minipage}
	}
	\subfigure[]{
	\begin{minipage}[t]{0.48\textwidth}
		\centering
		\includegraphics[width=\textwidth]{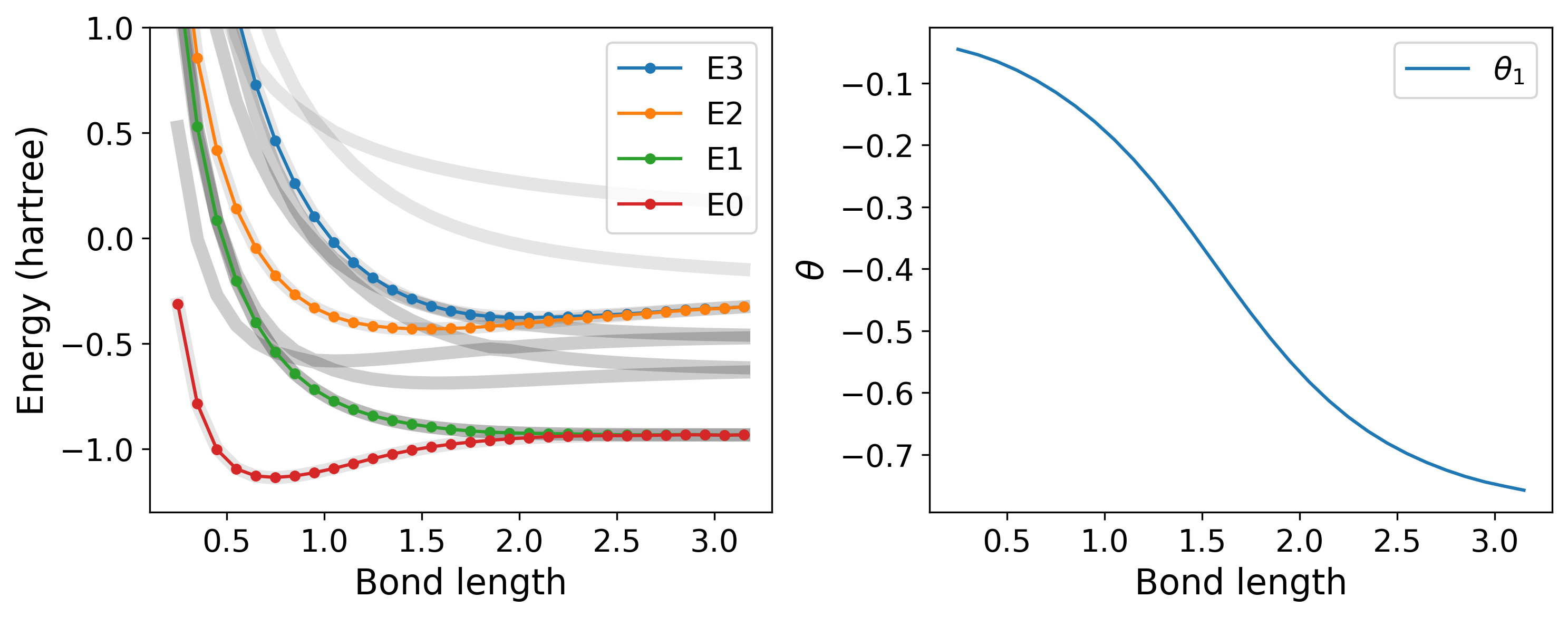}
		\end{minipage}
	}
	\caption{Landscape of H$_2$ in parameter space. (a)~Left: The black curve is corresponding to VQE, optimizing $\theta$ with respect to different bond length. The blue curve shows the process of Hamiltonian matching which optimize $\lambda$ with fixed $\theta$. The polylines with arrows in red, yellow and purple demonstarte the process of mutual gradient descent~(MGD) with different strategies(by controlling hyper parameters $N,T$) which converge to equilibrium point. Right: The polylines with arrows in grey shows the result of potential energy curve of H$_2$ with STO-3G basis sets from a randomly initialized point. The black one is optimized by a few steps of gradient descent to fix the error. (b)~Left: PEC calculated with different methods. Grey curves are exact diagonal results without fixing the electronic neutral condition. Colored curves are ground states and excited states that calculated with differential equations using STO-3G basis and UCC ansatz. Right: First component of $\boldsymbol{\theta}$ varies with bond length}
	\label{fig:h2}
\end{figure*}

\section{Numerical results}
\label{sec:result}
In this section, we will apply above methods in different systems including molecular hydrogen (H$_2$ and H$_4$) and Lithium Hydride. With those representative molecules, the results show that mutual gradient descent algorithm can search for equilibrium point in PEC in an enlarged parameter space and converge with only a few steps. Moreover, by solving the differential equation, we can calculate the PEC/PES with same accuracy as normal VQE does. This method can also avoid parameters to be trapped in local minima once the initial point is in global minima. Moreover, it is also compatible with existing VQE for calculating excited states. 
\begin{figure*}
	\subfigure[MGD optimization epoch]{
		\begin{minipage}[t]{0.3\textwidth}
			\centering
			\includegraphics[width=\textwidth]{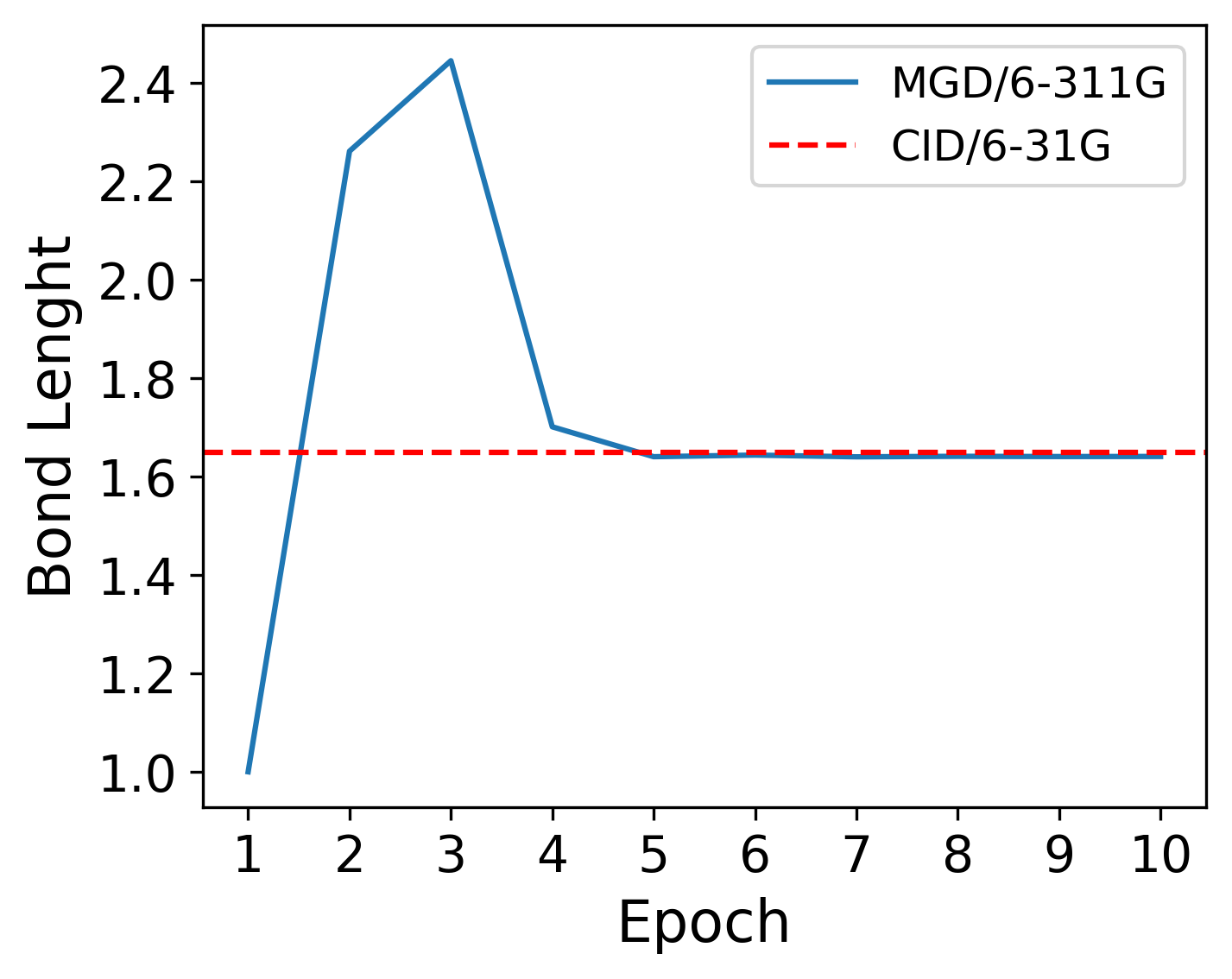}
		\end{minipage}
	}
	\subfigure[$\boldsymbol{\theta}$ flows with differential equations]{
		\begin{minipage}[t]{0.36\textwidth}
			\centering
			\includegraphics[width=\textwidth]{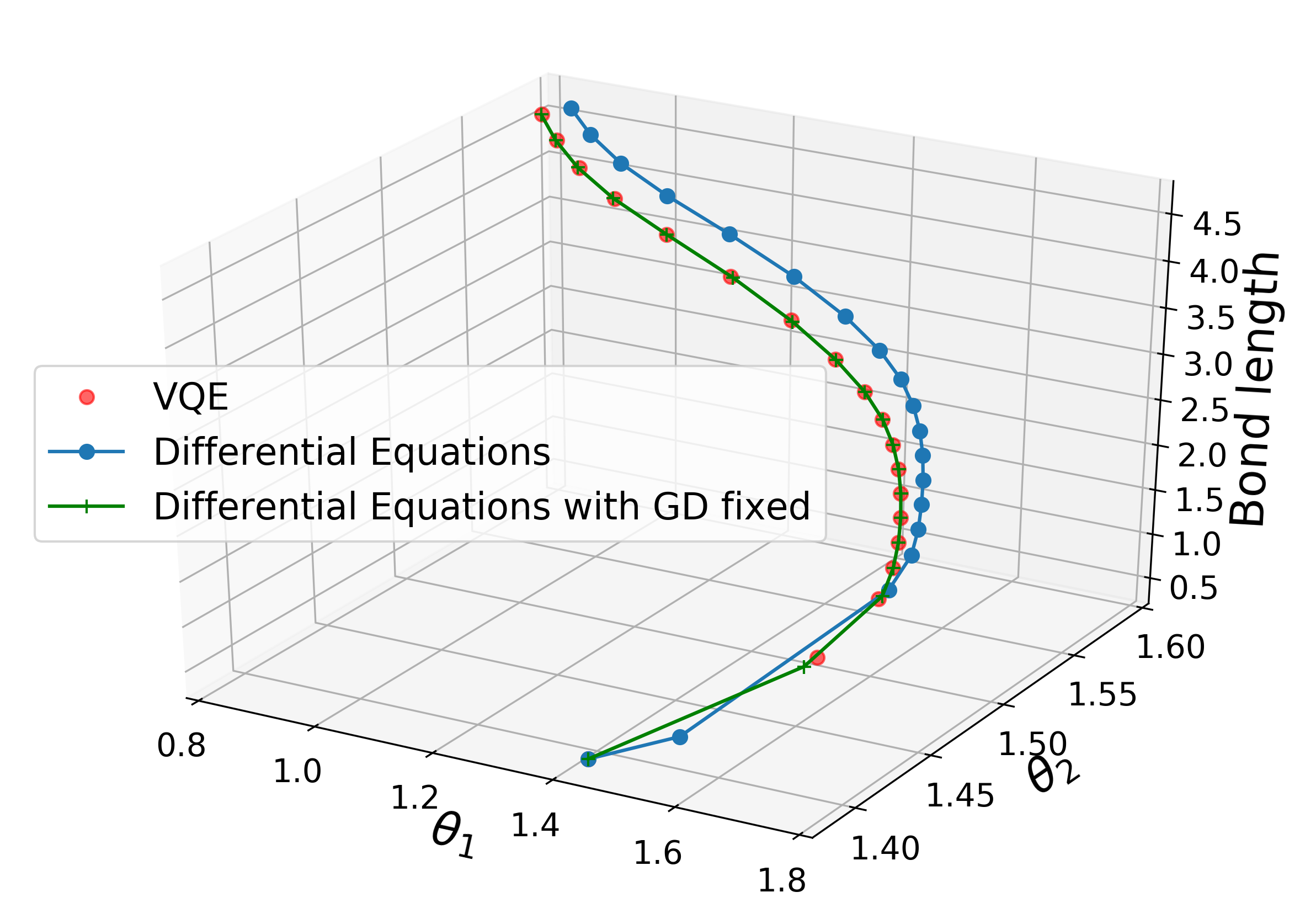}
		\end{minipage}
	}
	\subfigure[PEC of LiH]{
		\begin{minipage}[t]{0.28\textwidth}
			\centering
			\includegraphics[width=\textwidth]{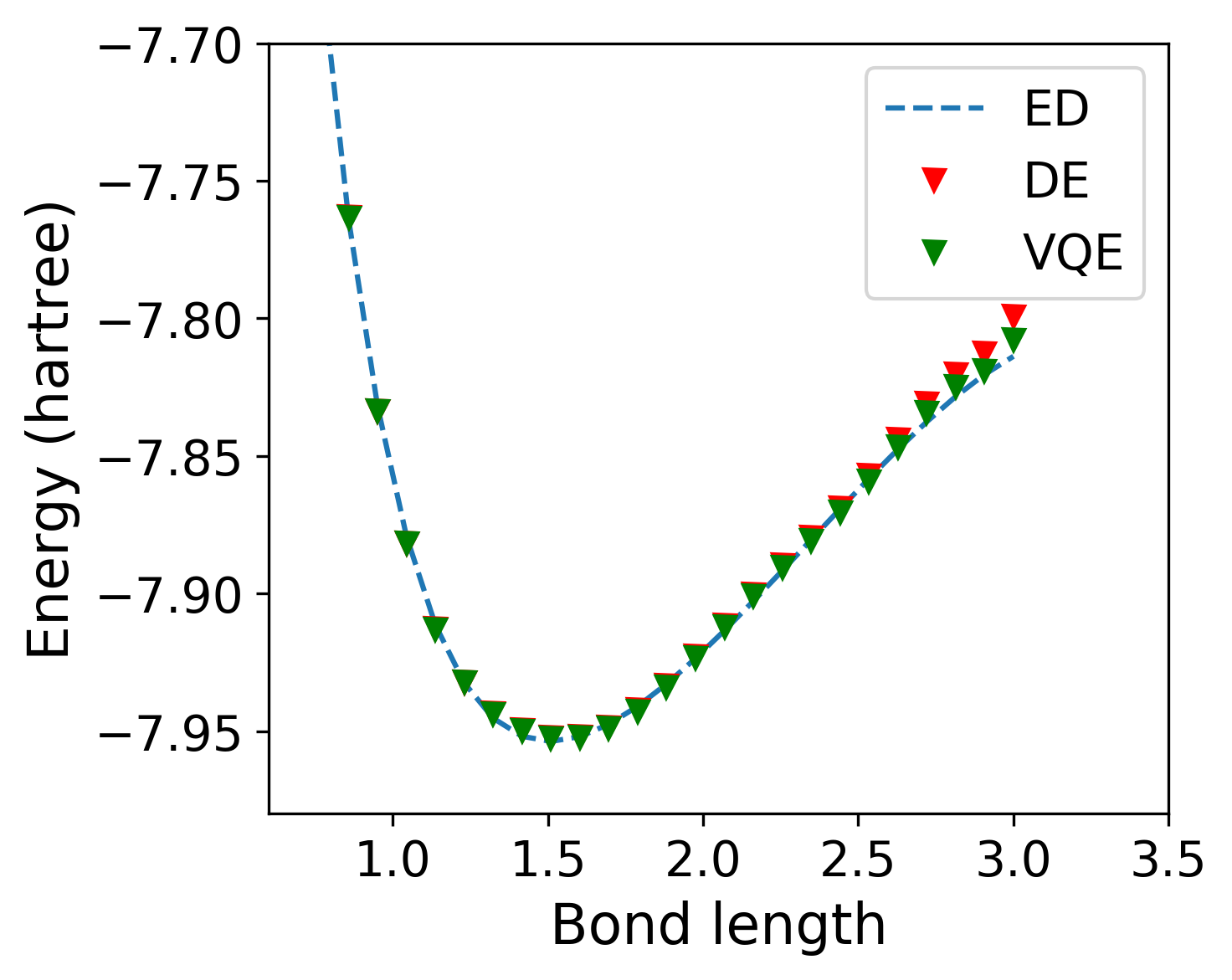}
		\end{minipage}
	}
	
	\caption{(a)Optimization process for finding equilibrium geometry with mutual gradient descent algorithm. Bond length is converge after only a few steps of iterative calculation. (b)The variations of parameters $\theta_1$ and $\theta_2$ as $\lambda$ changes calculated with differential equations. The red dots are VQE results and blue dots are results computed with differential equation. For omitting higher order terms in those equations, there are deviations with respect to VQE results. The green crosses are results by adding a few steps of gradient descent which is almost the same comparing to VQE. (c)PEC calculated with different methods,ED~(exact diagonal), DE~(differential equations), VQE~(variational quantum eigensolver).}
	\label{fig:lih}
\end{figure*}

\begin{table*}
	\centering
	\caption{Equilibrium geometry of H$_2$ and LiH}
	\begin{tabular} {*{5}{p{.2\textwidth}}}
		\hline \hline
		\toprule
		~&MGD&MGD&CID/6-31G&HF/6-31G\\ 
		\hline
		\midrule
		H$_2$~(a.u.)&0.744(STO-3G)&0.753(6-31G)&0.746&0.730\\
		LiH~(a.u.)&1.520(STO-6G)&1.641(6-311G)&1.649&1.636\\
		\hline \hline
		\bottomrule
	\end{tabular}
	\label{tbl:t1}
\end{table*}

\subsection{Molecular Hydrogen H$_2$}
We first use H$_2$ to illustrate basic properties and advantages of two algorithms for geometrical optimization and EPS calculations. We adopt both two and four qubits effective Hamiltonians obtained with OpenFermion package~\cite{OpenFermion} beforehand.

We adopt the unitary coupled cluster(UCC) ansatz with an unitary operator $U(\theta) = e^{-i\theta \sigma_0^x \sigma_1^y}$ acting on the Hartree-Fock reference state $\kets{01}$~\cite{YShen2017}. Bond lengths ranging from 0.2 a.u. to 2.85 a.u. was discretized uniformly into 50 points. Both gradient descent and scientific computing packages such as SciPy are available when using gradient-based methods to optimize the parameter $\theta$. Fig.~\ref{fig:h2} shows different processes, including traditional VQE and mutual gradient descent algorithm, of finding the balance point.

Traditional VQE requires to obtain the whole potential energy curve and then locate the minimum energy point while mutual gradient descent could converge to balance point with much fewer steps. To further understand the route of mutual gradient descent, we show the Hamiltonian matching results in the landscape. The blue line is the process of finding best matched bond length with different fixed $\theta$ which corresponds to find parent Hamiltonian with respect to a specific wavefunction. 

What impressed us is mutual gradient descent is following a path between Hamiltonian matching step(blue line) and wavefunction matching step(VQE). The predisposition of the curve is due to the strategy we choose which can further control the portion of consumption between quantum computing resource and classical computing resource~(see Fig.\ref{fig:h2}(a)). For comparison, more complicated variational ansatz~(4 qubits) and basis~(6-31g) are used in MGD. We use results from previous works of equilibrium geometry to compare our methods to traditional ones~(see Table.~\ref{tbl:t1})~\cite{DeFrees1979,DeFrees1982}.

We also apply differential equations which attempt to explicitly establish the dependence of optimized $\theta$ on $\lambda$. From our numerical results, we notice that potential surface curve(surface) can be obtained by solving those equations indeed. However, we also find that the deviation of the result becomes larger along the line of integration, which is due to finite step sizes
(see Fig.\ref{fig:h2}(b)). Those deviations can be further remedied with a few step of gradient descents.

\emph{Excited states.} Differential equations can also be used to calculate excited states. Combined with the idea of weighted SSVQE introduced in Ref.~\cite{Nakanishi2019}, which utilize the orthogonality between different eigenstates~(Hartree-Fork reference states, in our cases), we use a cost function as a weighted summation of energies,
\begin{equation}
\mathcal{L}_w(\theta) = \sum_{j=0}^k \omega_j \bras{\psi_j} U^{\dagger}(\theta)\mathcal{H}U(\theta)\kets{\psi_j}
\label{eq:ssvqe}
\end{equation}
instead of Eq.\eqref{eq:ene} in differential equations. Each state in Eq.\eqref{eq:ssvqe} required to be orthogonal with each other($\langle{\psi_i}|{\psi_j}\rangle=\delta_i^j$). The weight vector $\boldsymbol{w}$ can be any value in $(0,1)$ and should chosen $w_i < w_j$ when $i < j$. After using Eq.\eqref{eq:ssvqe} in differential equations, $\kets{\psi_j}$ will be the $j_{th}$ excited states. We use H$_2$ as examples and results can be seen in Fig.\ref{fig:h2}(b). Six orthogonal initial states are chosen $\kets{1100},\kets{1010},\kets{1001}$,$\kets{0110},\kets{0101},\kets{0011}$. As UCC operator keeps the particle number, those parameterized wavefunctions spans a subspace of electronic neutral states of H$_2$, which correspond to four energy curves in Fig.~\ref{fig:h2}(b), as some of them are degenerate.

\subsection{Lithium Hydride}
We now consider a more complicated molecule, LiH.  We first use three-qubit effective Hamiltonians~(six-qubit Hamiltonian will be used later) that construct with STO-6G and 6-311G basis sets for illustration of geometry optimization. Following Ref.~\cite{Higgott2019}, we use an UCC ansatz with two parameters, $
	U(\theta_1,\theta_2) = e^{-i\theta_2 \sigma_0^x \sigma_2^y} e^{-i\theta_1 \sigma_0^x \sigma_1^y}$,
and use $\kets{001}$ as Hartree-Fock reference state, more detail of UCC implementation can be seen in Appendix.\ref{append:ucc}. Bond lengths ranging from 0.2 a.u. to 3 a.u. are discretized uniformly into 30 points. 

In this case, we found the convergence of stationary point only needs a few steps. This is reasonable since it means parametrized wavefunction with different parameters are all in local minima with only slightly different bond length. This may not a general argument but it seems work well in all simple cases we study. Six qubits variational ansatz and 6-311g basis are used in MGD, too. And results from previous works of equilibrium geometry to compare our methods to traditional ones can be seen in Table.~\ref{tbl:t1}. The optimization process of differential equations for three-qubit Hamiltonian are shown in Fig.~\ref{fig:lih}(a).

We also use differential equations for calculating potential energy curve of LiH~(see Fig.~\ref{fig:lih}(c)). By canceling the error produced by finite step size with a few steps of gradient descent, the result is as good as VQE. For more accurate results, we increase the complexity of our quantum circuit. We use a 6 qubits UCC ansatz for calculating the potential energy curve with differential equations and VQE.

\begin{figure*}
	\subfigure[]{
	\begin{minipage}[t]{0.45\textwidth}
		\centering
		\includegraphics[width=\textwidth]{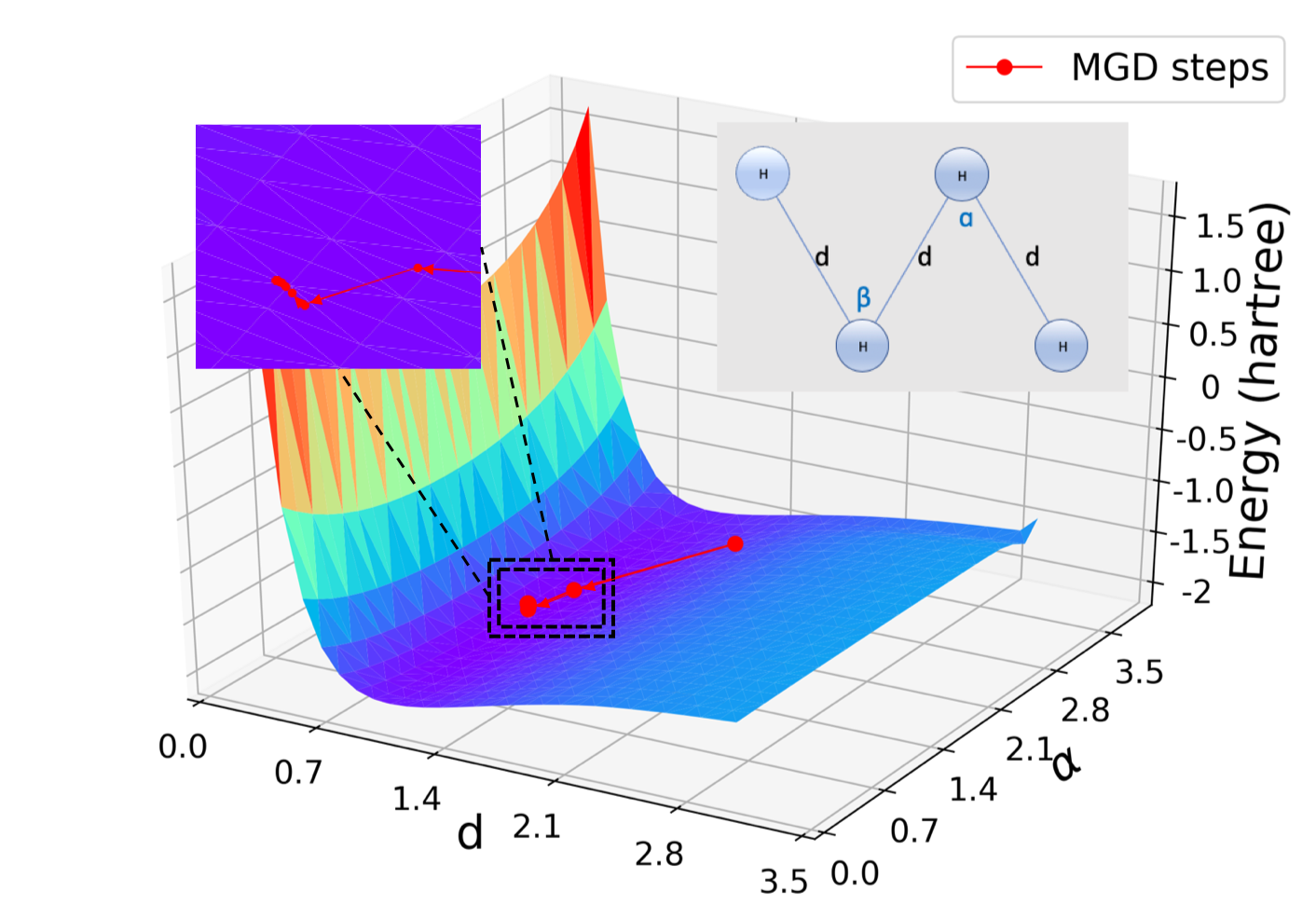}
	\end{minipage}
	}
		\subfigure[]{
	\begin{minipage}[t]{0.45\textwidth}
		\centering
		\includegraphics[width=\textwidth]{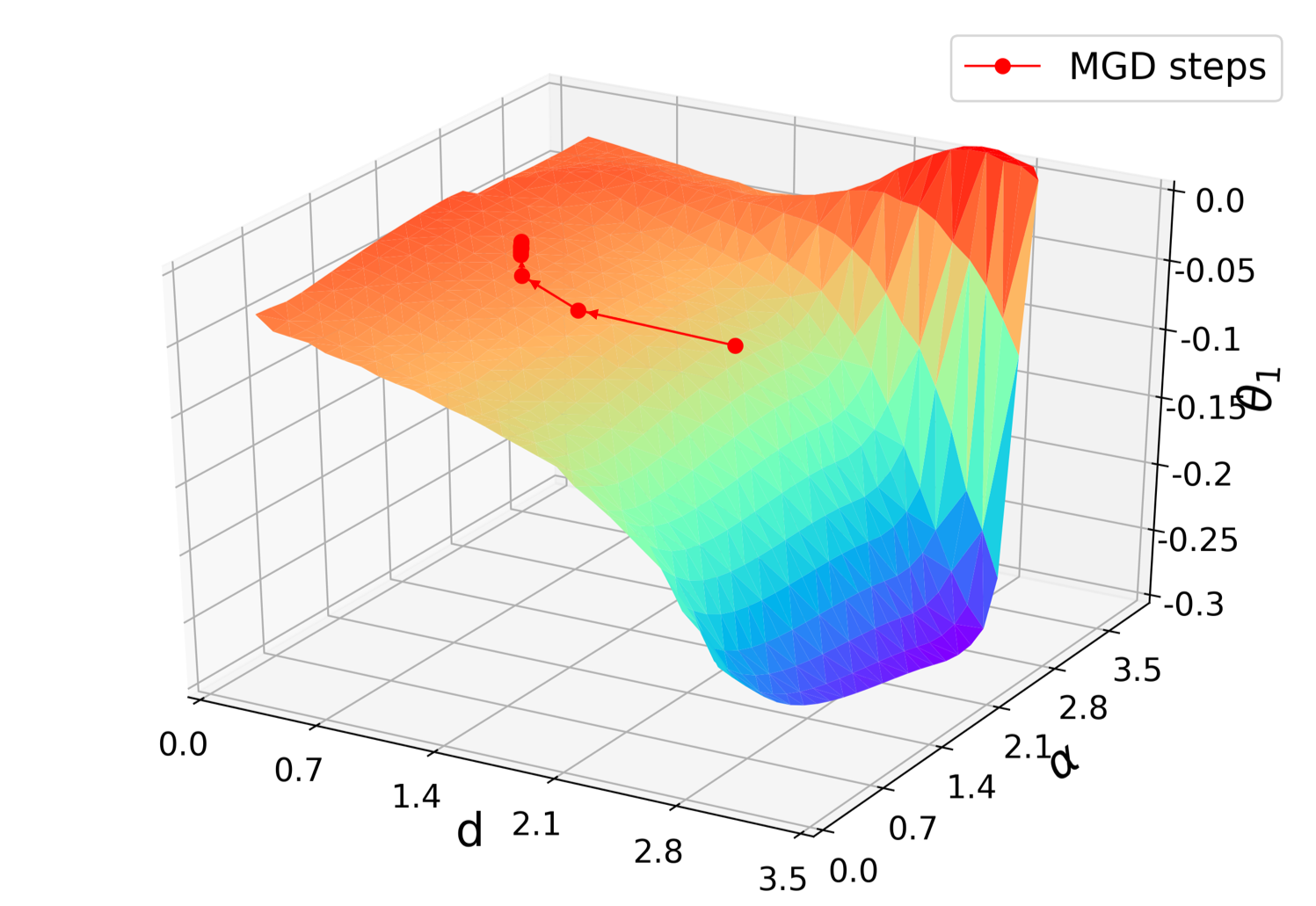}
		\end{minipage}
	}

	\caption{(a)~The process of finding equilibrium geometry with mutual gradient descent algorithm. The inset shows H$_4$ molecule model with four hydrogen atoms located on a broken line with equal distance $d$, angle $\beta = 60^\circ$, $d$ and angle $\alpha$ are two degrees of freedom.(b)~ First component of $\boldsymbol{\theta}$ varies with bond length $d$ and bond angle $\alpha$. }
	\label{fig:H4}
\end{figure*}

\subsection{H$_4$ molecule}
Molecules with many atoms are challenging in quantum chemistry, since the Hilbert space growths exponentially and the wavefunciton ansatz should be much more complicated. One protocol model can be multi-hydrogen systems such as hydrogen chain, which may be a benchmark for computational methods and also exhibits interesting physical phenomenons~\cite{Motta2019,Motta2017}. So we apply MGD to a H$_4$ molecule in order to exhibit the potential of this algorithm. 

For simplicity, we constraint this molecule in a way that only need 2 parameters to describe (see Fig.~\ref{fig:H4}). This molecule model is parametrized with bond length $d$ and bond angle $\alpha$. A more complicated unitary coupled clusters ansatz is used here and a 6 qubits quantum circuit is established for simulating it's eigenstates. The detail of UCC implementation is illustrate in Appendix.\ref{append:ucc}.

Bond lengths ranging from 0.3 a.u. to 3 a.u. are discretized uniformly into 24 points and bond angles are discretized uniformly into 25 points from $\pi/20$ to $25\pi/20$. 
The only difference between H$_4$ and previous examples is that the degrees of freedom become larger when $\boldsymbol{\lambda} = (d,\alpha)$. Since technically there is no difficult to compute the gradients with higher dimensional parameters, mutual gradient descent algorithm work well in this case too~(see Fig.\ref{fig:H4}(a)). MGD can find equilibrium geometry with a few steps in our molecule model. The parameters are continuously moving along the intrinsic degrees of freedom in parameter spaces~(see Fig.\ref{fig:H4}(b)).

\section{Summary}
\label{sec:summary}
To summary, we have proposed hybrid quantum-classical algorithms incorporated Hamiltonian derivative information for solving potential energy surface in quantum chemistry. Firstly, we have developed a method called mutual gradient descent algorithm, which have been shown to be efficient while finding equilibrium geometry under the context of VQE. MGD algorithm has successfully found the equilibrium bond length and bond angle of molecules like H$_2$, LiH and H$_4$ with only a few steps of iteration. We have developed differential equations incorporating Hamiltonian derivatives to calculate energy potential surface of molecules, which is also applicable for excited states. The paradigm of solving quantum chemistry in a combined Hamiltonian and wavefunction space on a quantum computer may inspire more practical quantum algorithms on near-term quantum devices. Lastly, we point out that while those quantum algorithms have been demonstrated for quantum chemistry, it is potential for applying them to solve quantum many-body problems with tunable parameters, such as quantum phase transitions. 
\begin{acknowledgments}
	This work was supported by the Key-Area Research and Development Program
	of GuangDong Province (Grant No. 2019B030330001), the National Key Research
	and Development Program of China (Grant No. 2016YFA0301800), the National Natural Science Foundation of China (Grants No. 91636218 and
	No. U1801661), the Key Project of Science and Technology of Guangzhou (Grant No. 201804020055).
\end{acknowledgments}

\appendix{
\section{Unitary coupled cluster ansatz}\label{append:ucc}
In quantum chemistry, unitary coupled cluster (UCC) ansatz is used widely for a parametrization of wavefunction due to its representation power. The variational wavefunction can be written as
\begin{equation}
	\kets{\psi(\boldsymbol{\theta})} = e^{T-T^\dagger} \kets{R},
\end{equation}
where $T = T_1+T_2$ with 
\begin{eqnarray}\label{eq:uccsd}
&&T_1=\sum_{pq} \theta^p_q c_p^\dagger c_q \nonumber \\
&&T_2=\sum_{pqrs}\theta^{pq}_{rs}c_p^\dagger c_q^\dagger c_r c_s.  
\end{eqnarray}

$T_1$ and $T_2$ represent single particle excitations and double particle excitations, respectively, and $c$ and $c^\dagger$ are fermionic operators. The ansatz we used here also called unitary coupled cluster with single and double excitations~(UCCSD).

This expression can not be implemented on quantum devices directly. Wigner transformation (or Bravyi-Kitaev transformation) and trotterization is needed before optimization. Trotter steps can be a single step or more based on required accuracy, the expression for n Trotter step is

\begin{equation}
\kets{\psi(\boldsymbol{\theta})}=\prod_{k=1}^{n}\prod_{pq}e^{\hat{t}^k_{pq}}\prod_{pqrs}e^{\hat{t}^k_{pqrs}}\kets{R}.
\end{equation}	
where $\hat{t}^k_{pq}=\theta^{p,k}_q(c_p^\dagger c_q-c_q^\dagger c_p)$ and
$\hat{t}^k_{pqrs}=\theta^{pq,k}_{rs}(c_p^\dagger c_q^\dagger c_r c_s-c_s^\dagger c_r^\dagger c_p c_q)$
}

%
\end{document}